\documentclass[11pt,a4paper]{article}

\usepackage{fullpage}
\usepackage{amsmath}
\usepackage{amssymb}
\usepackage{mathrsfs}
\usepackage{graphicx}
\usepackage{psfrag}
\usepackage[all,poly,curve,knot,cmtip]{xy}
\usepackage{citesort}

\newcommand {\be} {\begin{eqnarray*}}
\newcommand {\ee} {\end{eqnarray*}}
\newcommand {\bea} {\begin{eqnarray}}
\newcommand {\eea} {\end{eqnarray}}

\newcommand{\bm}[1]{\boldsymbol{#1}}

\newcommand{\ave}[1]{\langle {#1} \rangle}

\newcommand{\pdiff}[2]{\frac{\partial{#1}}{\partial{#2}}}

\def\M{\mathcal{M}}
\def\ez{\,{}^o\! e}

\def\f{\frac}

\title{First-order action and Euclidean quantum gravity}
\author{\textbf{Tom$\acute{\mbox{a}}\check{\mbox{s}}$
Liko}\footnote{On leave from Dept. Physics and Physical Oceanography,
Memorial University of Newfoundland, St. John's, NL, Canada A1B 3X7.
Electronic mail: liko@gravity.psu.edu} ~ and
\textbf{David Sloan}\footnote{Electronic mail: sloan@gravity.psu.edu}\\
\\{\small \it Institute for Gravitation and the Cosmos}\\
{\small \it Pennsylvania State University}\\
{\small \it University Park, Pennsylvania 16802, U.S.A.}}

\begin{document}

\maketitle




\begin{abstract}

We show that the on-shell path integral for asymptotically flat Euclidean spacetimes
can be given in the first-order formulation of general relativity, without assuming
the boundary to be isometrically embedded in Euclidean space and without adding infinite
counter-terms.  For illustrative examples of our approach, we evaluate the first-order
action for the four-dimensional Euclidean Schwarzschild and NUT-charged spacetimes to
derive the corresponding on-shell partition functions, and show that the correct
thermodynamic quantities for the solutions are reproduced.

\end{abstract}

\hspace{0.35cm}{\small \textbf{PACS}: 04.70.Bw; 04.20.Cv}



\section{Introduction}

To study the thermodynamics of any system from a quantum-mechanical point of view, the quantity
of interest is the partition function.  In the operator representation, this is
\bea
\mathcal{Z} = \mbox{Tr}\left[\mbox{exp}\left(-\beta\hat{H}[\phi]\right)\right] \, ,
\label{operatorpartition}
\eea
for a system of fields $\phi$ at finite temperature $T=1/\beta$ with Hamiltonian
$\hat{H}[\phi]$.  In the path integral representation (\ref{operatorpartition}) is
equivalent to the expression \cite{hawking}
\bea
\mathcal{Z} = \int\mathcal{D}[\phi]\mbox{exp}\left(-\tilde{I}[\phi]\right) \, ,
\label{epath1}
\eea
where $\tilde{I}[\phi]$ is the classical action.

Evaluation of $\mathcal{Z}$, however, is quite difficult in practice: the integration and the
measure $\mathcal{D}[\phi]$ are difficult to construct.  The problem is that the integration
is over all fields $[\phi]$, not just the classical fields $\phi_{0}$ that satisfy the equations
of motion $\delta\tilde{I}[\phi_{0}]=0$.  For physical applications, however, it is reasonable
to expect that the dominant contributions to the partition function will come from fields that
are close to the classical fields (i.e. the stationary-phase approximation).  So for a field
$\phi=\phi_{0}+\delta\phi$ the action can be expanded in a Taylor series such that
\bea
\tilde{I}[\phi_{0}+\delta\phi] = \tilde{I}[\phi_{0}] + \delta\tilde{I}[\phi_{0},\delta\phi]
                                 + \delta^{2}\tilde{I}[\phi_{0},\delta\phi] + \ldots \; .
\eea
For a generic field $\phi$ the first term $\tilde{I}[\phi_{0}]$ is assumed to be finite,
the linear term $\delta\tilde{I}$ is assumed to vanish, while the quadratic term
$\delta^{2}\tilde{I}$ is assumed to be positive-definite \cite{grumcn}.  If these
three properties hold, then the \emph{on-shell} partition function may be approximated to the
expression
\bea
\mathcal{Z} = \mbox{exp}\left(-\tilde{I}[\phi_{0}]\right)
\label{epath2}
\eea
This form of the partition function is known as the stationary-phase approximation to the
path integral (\ref{epath1}).  The standard quantities of thermodynamics can then
be calculated.  In particular, the average energy $\ave{E}$ and entropy $S$ are given by
\bea
\ave{E} = -\pdiff{\ln\mathcal{Z}}{\beta}
\quad
\mbox{and}
\quad
S = \beta\ave{E} + \ln\mathcal{Z} \; .
\label{eands}
\eea
The physical meaning of the energy may differ based on the boundary conditions that are used,
i.e. holding the pressure or volume constant.

For gravitational objects that satisfy the Einstein equations, the partition function
of interest is evaluated for the Einstein-Hilbert action on a $D$-dimensional manifold
$\mathcal{M}$ in a bounded region \cite{york1,gibhaw1,hawking}:
\bea
I[g] = \frac{1}{2\kappa}\int_{\mathcal{M}}Rd^{D}V
       + \frac{1}{\kappa}\oint_{\partial\mathcal{M}}Kd^{D-1}V \, ,
\label{ehaction}
\eea
where $\kappa=8\pi$ (with $G_{D}=1$), $\partial\mathcal{M}$ is the boundary of $\mathcal{M}$,
$R$ is the Ricci scalar of the spacetime metric $g$ and $K$ is the trace of the extrinsic
curvature of the boundary $\partial\mathcal{M}$, $d^{D}V$ is the volume element determined
by $g$, and $d^{D-1}V$ is the volume element determined by the induced metric $h$ on
$\partial\mathcal{M}$.  The surface term in (\ref{ehaction}) can be understood to arise in
the action principle for general relativity because the Lagrangian density depends on the
second derivatives of the metric.  As a result the first derivatives of $g$ in addition to
$g$ itself must be held fixed on $\partial\mathcal{M}$.  This is in contrast to the usual
form of variational principles for which only the field itself is held fixed.  If $\mathcal{M}$
is spatially compact then a well defined variational principle for the action (\ref{ehaction})
exists.  In this case (and in this case only), the partition function is well defined, and
the stationary-phase approximation gives
\bea
\mathcal{Z} = \mbox{exp}\left(-\tilde{I}[g_{0}]\right)
\label{epath3}
\eea
From here one can then calculate the average energy and entropy of the spacetime with metric
$g_{0}$.

The above prescription for finding the thermodynamic properties of gravitational objects works
well if $\mathcal{M}$ is spatially compact, but contains the seeds of many problems if
$\mathcal{M}$ is asymptotically flat: in the latter case the action (\ref{ehaction}) is infinite,
even in the flat limit.  This would imply that the partition function (\ref{epath2})
evaluated on asymptotically flat Euclidean spacetimes is ill-defined.  The solution (for Lorentzian
spacetimes) is to isometrically embed the boundary manifold $(\partial\mathcal{M},h)$ in Minkowski
spacetime, calculate the extrinsic curvature $K_{0}$ of $\partial\mathcal{M}$ defined by the
Minkowski metric, and subtract the resulting quantity from the boundary integral in (\ref{ehaction}).
Thus the action \cite{gibhaw1}
\bea
I[g] = \frac{1}{2\kappa}\int_{\mathcal{M}}Rd^{D}V
       + \frac{1}{\kappa}\oint_{\partial\mathcal{M}}(K - K_{0})d^{D-1}V
\label{countertermaction}
\eea
gives a well defined action principle for asymptotically flat spacetimes.  One may then proceed
to find the partition function for asymptotically flat spacetimes such as the Schwarzschild or
Kerr solutions, and hence the thermodynamic quantities of interest.  The problem with the
infinite subtraction in the action (\ref{countertermaction}) is that the embedding scheme does not
work for generic spacetimes in dimensions $D\geq4$ \cite{hawking,manmar,aes,ashslo}.  This is because
the Gauss-Codazzi equation (in $D>3$ dimensions) cannot be generically solved for the extrinsic
curvature.  In addition, the infinite subtraction method crucially depends on the topology of the
system under consideration.

Since the original work of Gibbons and Hawking \cite{gibhaw1} many proposals for counter-terms have
been proposed.  See e.g. \cite{lau,mann,kls,manmar,mmv,ams,grumcn}.  These counter-term methods resolve
the limitations of the infinite subtraction method, and have led to an understanding of the thermodyanmics
of a number of solutions that was not possible before.  Physically, however, it is desirable to employ a
framework that generically produces finite quantities \emph{without the need of adding any counter-terms}.
As was recently shown \cite{aes,ashslo}, the first-order formulation of general relativity based on
orthonormal co-frames and Lorentz connections as independent fields does provide such a framework for
asymptotically flat spacetimes in $D\geq4$ dimensions.  The purpose of this paper is to show that a
partition function can be given for asymptotically flat Euclidean manifolds in this framework.

\section{Finiteness of the first-order action}

We proceed with our discussion of the properties of the first-order action parallel to that in
\cite{ashslo}.  The Cartesian coordinates $x^a$ of the \emph{flat metric}
$g^o_{ab}=\delta_{IJ}{}^{0}e_{a}^{\phantom{a}I}{}^{0}e_{b}^{\phantom{a}J}$ and the associated radial
coordinates $(r,\Phi^i)$ (with $r^{2}=\delta_{ab}x^{a}x^{b}$ and $\Phi^{i}$ the standard angular coordinates
on hyperboloids defined by $r=\mbox{constant}$) will be used in asymptotic expansions.  Detailed analysis
shows that to define the angular momentum one needs $e_{a}^{\phantom{a}I}$ to admit
an expansion to order $D-2$. Therefore, we will assume that $e_{a}^{\phantom{a}I}$ can be expanded as:
\bea
e = {}^{0}\!{e}(\Phi) + \f{{}^{D-3}e (\Phi)}{r^{D-3}}
+ \f{{}^{D-2}e(\Phi)}{r^{D-2}} + \mathcal{O}(r^{D-1}) \, ,
\label{ed}
\eea
with a reflection symmetric ${}^{D-3}\!e(\Phi)$.  Here and in what follows
\bea
r_a = \partial_a r \quad {\rm and}\quad  r^I = \eta^{IJ}{}^{0}\! e_{\phantom{a}J}^{a}r_{a} \; .
\eea
To appropriate leading orders, $A_a^{\phantom{a}IJ}$ can be required to be compatible with $e_{a}^{\phantom{a}I}$
on the boundary $\partial \M$ of $\M$.  This leads us to require that $A_a^{\phantom{a}IJ}$ is asymptotically
of order
$D-1$,
\bea
A = {}^{0}\!{A}(\Phi) + \f{{}^{1}\!{A}(\Phi)}{r} + \ldots + \f{{}^{D-1}\!{A}(\Phi)}{r^{D-1}} + \mathcal{O}(r^{D-1}) \; .
\label{Ad}
\eea
Compatibility of $A$ with $e$ and flatness of $\ez$ enables us to set ${}^{0}\!{A} =\ldots= {}^{D-3}\!{A} = 0$
and express ${}^{D-2}\!A$ as
\bea
{}^{D-2}\!{A_{a}^{\phantom{a}IJ}}(\Phi) = 2r^{D-2}\, \partial^{[J}\,\left(r^{3-D}\,\, {}^{1}\!e_{a}^{\phantom{a}I]} \right) \; .
\label{Ad2}
\eea

We consider a $D$-dimensional \emph{Euclidean} manifold bounded by two spacelike Cauchy surfaces,
$M_{1}$ and $M_{2}$, which are asymptotically related by a time translation.  In the first-order
formulation of general relativity the action is given by (see e.g. \cite{ashslo})
\bea
\tilde{I}[e,A] = \frac{1}{2\kappa}\int_{\mathcal{M}}\Sigma_{IJ} \wedge \Omega^{IJ}
                 - \frac{1}{2\kappa}\oint_{\partial\mathcal{M}}\Sigma_{IJ} \wedge A^{IJ} \; .
\label{action2}
\eea
This action depends on the co-frame $e^{I}$ and the $SO(D)$ connection $A_{\phantom{a}J}^{I}$.  The
co-frame determines the metric $g_{ab}=\delta_{IJ}e_{a}^{\phantom{a}I} \otimes e_{b}^{\phantom{a}J}$,
$(D-2)$-form
$\Sigma_{IJ}=[1/(D-2)!]\epsilon_{IJK_{1}\ldots K_{D-2}}e^{K_{1}} \wedge \cdots \wedge e^{K_{D-2}}$
and spacetime volume form $\bm{\epsilon}=e^{0} \wedge \cdots \wedge e^{D-1}$, where
$\epsilon_{I_{1}\ldots I_{D}}$ is the totally antisymmetric Levi-Civita tensor.  The connection
determines the curvature two-form
\bea
\Omega_{\phantom{a}J}^{I} = dA_{\phantom{a}J}^{I}
+A_{\phantom{a}K}^{I} \wedge A_{\phantom{a}J}^{K}
= \frac{1}{2}R_{\phantom{a}JKL}^{I}e^{K} \wedge e^{L} \, ,
\eea
with $R_{\phantom{a}JKL}^{I}$ as the Riemann tensor.  Internal indices $I,J,\ldots\in\{0,\ldots,D-1\}$
are raised and lowered using the flat metric $\delta_{IJ}=\mbox{diag}(1,\ldots,1)$.

Now, as was explicitely shown in \cite{aes} and \cite{ashslo}, ``power counting'' arguments are sufficient
to show that $\tilde{I}$ is finite and that $\delta \tilde{I}=0$ on shell.  Thus the limitations of the
infinite subtraction method discussed in the Introduction are resolved as a consequence of the natural
boundary conditions for asymptotic flatness in the first-order formalism, without having to add counter-terms.
Therefore the on-shell partition function in the first-order framework is
\bea
\mathcal{Z} = \mbox{exp}\left(-\tilde{I}[e_{0},A_{0}]\right) \; .
\label{epath5}
\eea
This is precisely the zero-loop contribution to the path integral (\ref{epath2}).  The partition function
(\ref{epath5}) is well defined because of two important features of the first-order action:
\begin{enumerate}
\item
It does not require, or make \emph{any} reference to, the embedding of $\partial\mathcal{M}$ in Euclidean
space;
\item
It is finite for asymptotically flat Euclidean manifolds without any counter-terms.
\end{enumerate}
These key properties are important for Euclidean quantum gravity techniques to be applicable to
generic spacetimes in dimensions $D\geq4$, regardless of topology.

\section{Correspondence between the first-order and second-order boundary terms}

An important feature of the boundary term in the action (\ref{action2}) is its inequivalence to the GHY
boundary term in the action (\ref{countertermaction}) for generic spacetimes with boundaries.  This can
be understood in the following way.  First, fix an internal gauge once and for all such that
$n_{a}=n_{I}e_{a}^{\phantom{a}I}$ is the unit normal to
$\partial\mathcal{M}\cong M_{1}\cup M_{2}\cup\tau_{\infty}$, and that $\partial_{a}n_{I}=0$ on $M_{1}$
and $M_{2}$.  Now we note that the first-order boundary term, in components, can be simplified to
\bea
\oint_{\partial\mathcal{M}}\Sigma_{IJ} \wedge A^{IJ}
= 2\oint_{\partial\mathcal{M}}\bm{\varepsilon}e^{aI}A_{aI}^{\phantom{aI}J}n_{J} \, ,
\label{components}
\eea
with $\bm{\varepsilon}$ the volume form on $\partial\mathcal{M}$ (for useful identities that make this
proof simple, see Section 2.3.1 of \cite{ashlew}).  Now we note that the extrinsic curvature, in general,
is given by
\bea
K_{ab} &=& h_{a}^{\phantom{a}c}h_{b}^{\phantom{a}d}\nabla_{c}n_{d}\nonumber\\
       &=& h_{a}^{\phantom{a}c}h_{b}^{\phantom{a}d}(\partial_{c}n_{I}
           + A_{cI}^{\phantom{aI}J}n_{J})e_{d}^{\phantom{a}I} \, ,
\eea
where we used the relation $n_{a}=n_{I}e_{a}^{\phantom{a}I}$ and the compatability of $e_{a}^{\phantom{a}I}$.
Taking the trace of the extrinsic curvature therefore gives the identity
\bea
e^{aI}A_{aI}^{\phantom{aI}J}n_{J} = K - e^{aI}\partial_{a}n_{I} \; .
\label{actionidentity}
\eea
It follows that on $\partial\mathcal{M}$ the first-order boundary term is given by
\bea
\oint_{\partial\mathcal{M}}\Sigma_{IJ} \wedge A^{IJ}
= 2\oint_{\partial\mathcal{M}}\bm{\varepsilon}(K - K_{0}^{\prime}) \, ,
\label{general}
\eea
where we defined the quantity $K_{0}^{\prime}=e^{aI}\partial_{a}n_{I}$.  From this correspondence we
can make the following two conclusions:
\begin{enumerate}
\item
On $M_{1}$ and $M_{2}$ we have that $\Sigma_{IJ} \wedge A^{IJ}=2\bm{\varepsilon}K$, which is
precisely the GHY term.
\item
On $\tau_{\infty}$ the first-order boundary term will coincide with the GHY term in (\ref{countertermaction})
if and only if there exists an isometric embedding of the boundary $\tau_{\infty}$ in Euclidean space (i.e.
where $e={}^{0}\!{e}(\Phi)$).
\end{enumerate}

To summarize, we can recover a second-order action by enforcing compatability between co-frame and connection
in the first-order action.  The resulting second-order action reproduces \emph{exactly} the GHY boundary term
if the boundary can be isometrically embedded in Euclidean space, as is required in the GHY prescription.
By contrast, no such emdedding is required in the first-order framework.  This important
feature will be illustrated in the next section, where we will reproduce the correct thermodynamical quantities
for NUT-charged spacetimes. 

We conclude this section by commenting on the gauge invariance of the first-order boundary term in (\ref{action2}).
At first glance it may seem that this boundary term breaks gauge invariance because of the explicit dependence of
the connection in the integrand.  However, this is not the case.  In particular, the gauge invariance on $M_{1}$
and $M_{2}$ follows trivially from the preceeding discussion: on $M_{1}$ and $M_{2}$ we have that
$\Sigma_{IJ} \wedge A^{IJ}=2\bm{\varepsilon}K$ which is manifestly gauge invariant.  On the other hand, on
 $\tau_{\infty}$ the co-frame $e$ tends to ${}^{0}e$ while permissible gauge transformations of $A$ tend to
identity (i.e. under infinitesimal gauge transformations).  Therefore our boundary term is also gauge invariant
on $\tau_{\infty}$.

\section{Partition functions and thermodynamics}

As illustrative examples of our formalism, we will now derive the partition functions (\ref{epath5}) and
hence the thermodynamic quantities in (\ref{eands}) for specific solutions to the Einstein equations.
For concreteness, we will evaluate the first-order action (\ref{action2}) for the Euclidean Schwarzschild,
Taub-NUT and Taub-bolt spacetimes in four dimensions.  All three examples are vacuum solutions, so that the
corresponding bulk actions do not contribute.

To evaluate the boundary terms, the standard prescription is to evaluate seperately the contributions from
the inner and outer boundaries by calculating the integrals on constant-$r$ hypersurfaces and taking the
limits as $r$ goes to the horizon and to infinity; for all three examples the contribution from the inner
limit is zero.  Therefore in what follows we will only provide details of the contribution at $\tau_{\infty}$.
In the first-order formalism, the calculation of $\tau_{\infty}$'s contribution amounts to calculating the
${}^{2}A$ contribution to the boundary integral, which can be obtained by expanding the co-frame in powers
of $r^{-1}$ and substituting the ${}^{1}e$ term in equation (\ref{Ad2}).

\subsection{Schwarzschild solution}

The metric for four-dimensional Schwarzschild spacetime with Euclidean time $\tau$ has line element
\cite{gibhaw2}
\bea
ds^{2} = f(r)d\tau^{2} + \frac{dr^{2}}{f(r)} + r^{2}(d\theta^{2} + \sin^{2}\theta d\phi^{2}) \, ,
\label{eschwarzschild}
\eea
with $f(r)=1-2M/r$ and $M$ the mass of the source.  Regularity of the metric at the point singularity $r=2M$
requires that $\tau$ have a period $\beta=8\pi M$.

A suitable tetrad of co-frames for this spacetime is given by
\bea
e^{0} = \sqrt{f}d\tau \, ,
\quad
e^{1} = \frac{1}{\sqrt{f}}dr \, ,
\quad
e^{2} = rd\theta \, ,
\quad
e^{3} = r\sin\theta d\phi \; .
\eea
Expanding this tetrad in powers of $r^{-1}$, we find that the only non-zero ${}^{1}e$ components are
\bea
{}^{1}e_{0}^{\phantom{a}0} = -M \, ,
\quad
{}^{1}e_{1}^{\phantom{a}1} = M \, ,
\eea
and substituting these into (\ref{Ad2}) gives
\bea
{}^{2}A_{0}^{\phantom{a}01} = M \, ,
\quad
{}^{2}A_{1}^{\phantom{a}01} = -M \; .
\eea
We therefore find that the Euclidean action is given by
\bea
\tilde{I} &=& \frac{1}{\kappa}\oint_{\tau_{\infty}}{}^{0}e_{2}^{\phantom{a}2}\,{}^{0}e_{3}^{\phantom{a}3}\,\frac{{}^{2}A_{0}^{\phantom{a}01}}{r^{2}}\partial_{1}r\nonumber\\
          &=& \frac{\beta^{2}}{16\pi} \; .
\eea
Substituting this in (\ref{epath5}) then gives the partition function
\bea
\mathcal{Z} = \mbox{exp}\left(-\frac{\beta^{2}}{16\pi}\right) \; .
\eea
The thermodynamic quantities can now be calculated.  From (\ref{eands}) we have that
\bea
\ave{E} = M
\quad
\mbox{and}
\quad
S = 4\pi M^{2} \; .
\eea
Replacing $M=r_{+}/2$ and noting that the surface area $A$ of the horizon is $A=4\pi r_{+}^{2}$
we get $S=A/4$ as should have been expected.

\subsection{NUT-charged solutions}

The metric for four-dimensional Taub-NUT spacetime with Euclidean time $\tau$ has line element
\cite{gibhaw2} (see also \cite{ams})
\bea
ds^{2} = V(r)\left[d\tau + 2N\cos\theta d\phi\right]^{2} + \frac{dr^{2}}{V(r)}
         + (r^{2} - N^{2})(d\theta^{2} + \sin^{2}\theta d\phi^{2}) \, ,
\label{etaubnut}
\eea
with $V(r)=(r^{2}-2Mr+N^{2})/(r^{2}-N^{2})$ and $N$ the NUT parameter.  Regularity of the metric
requires that $\tau$ have a period $\beta=8\pi N$.

A suitable tetrad of co-frames for this spacetime is given by
\bea
e^{0} &=& \sqrt{V}d\tau + 2\sqrt{V}N\cos\theta d\phi \, ,
\quad
e^{1} = \frac{1}{\sqrt{V}}dr \, ,
\quad
e^{2} = \sqrt{r^{2} - N^{2}}d\theta \, ,\nonumber\\
e^{3} &=& \sqrt{r^{2} - N^{2}}\sin\theta d\phi \; .
\label{tnuttetrad}
\eea
Expanding this tetrad in powers of $r^{-1}$, we find that the non-zero ${}^{1}e$ components are
\bea
{}^{1}e_{0}^{\phantom{a}0} = -M \, ,
\quad
{}^{1}e_{3}^{\phantom{a}0} = 2N\cos\theta \, ,
\quad
{}^{1}e_{1}^{\phantom{a}1} = M \, ,
\eea
and substituting these into (\ref{Ad2}) gives
\bea
{}^{2}A_{0}^{\phantom{a}01} = M \, ,
\quad
{}^{2}A_{1}^{\phantom{a}01} = -M \, ,
\quad
{}^{2}A_{3}^{\phantom{a}01} = -2N\cos\theta \, ,
\quad
{}^{2}A_{3}^{\phantom{a}02} = -2N\cos\theta \; .
\eea
The Euclidean action for the Taub-NUT spacetime is therefore given by
\bea
\tilde{I} &=& \frac{1}{\kappa}\oint_{\tau_{\infty}}{}^{0}e_{2}^{\phantom{a}2}\,{}^{0}e_{3}^{\phantom{a}3}\,\frac{{}^{2}A_{0}^{\phantom{a}01}}{r^{2}}\partial_{1}r\nonumber\\
          &=& 4\pi MN \; .
\eea
Substituting this in (\ref{epath5}) then gives the partition function
\bea
\mathcal{Z} = \mbox{exp}\left(-4\pi MN\right) \; .
\label{parfun}
\eea
This result agrees exactly with the action that was computed by Astefanesei \emph{et al} \cite{ams} using the
Mann-Marolf counter-term method.  The thermodynamic quantities can now be calculated.  In particular, substituting
$M=N$ into (\ref{parfun}) we find the average energy and entropy for the ``NUT'' charge given by
\bea
\ave{E} = N
\quad
\mbox{and}
\quad
S = 4\pi N^{2} \, ,
\eea
while substituting $M=5N/4$ into (\ref{parfun}) we find the average energy and entropy for the ``bolt'' charge
given by
\bea
\ave{E} = \frac{5N}{4}
\quad
\mbox{and}
\quad
S = 5\pi N^{2} \; .
\eea
Therefore we find agreement with previous results due to Mann \cite{mann} and Astefanesei \emph{et al} \cite{ams}.

\section{Discussion}

Since the pioneering work of Gibbons and Hawking \cite{gibhaw1} on the Euclidean path integral methods of black-hole
thermodynamics, calculations of partition functions have been almost exclusively done in the second-order formalism.
In this formalism, however, the semiclassical approximation of considering small perturbations around a classical
asymptotically flat solution faces two obstacles: (1) the action is not finite, even on-shell; and (2) the linear
term in the perturbation need not vanish.  These two problems can both be solved by adding counter-terms
to the action \cite{lau,mann,kls,manmar,mmv,ams,grumcn}, but these are model-specific \emph{post hoc} additions.

By contrast, the first-order action is finite on-shell under the natural boundary conditions arising from asymptotic
flatness \cite{aes,ashslo}, and does not require the boundary to be isometrically embedded in Euclidean space.  We
have shown here that the Euclidean path integral in the first-order formalism  yields a well defined partition
function without the need of adding by hand any counter-terms.

The first-order action, when evaluated on the Euclidean Schwarzschild, Taub-NUT and Taub-bolt solutions, agree
exactly with the counter-term methods in the second-order formalism.  In turn, the corresponding partition functions
for these solutions in the first-order and second-order frameworks are identical.  The simplified manner by which
this was achieved in the first-order formalism relative to the second-order formalism suggests that this provides
a more solid basis for quantum theory.  

\section*{Acknowledgements}

We wish to thank Abhay Ashtekar and Andrew Randono for discussions.  We also thank Abhay Ashtekar for suggesting
the initial direction for the work presented here, and for commenting on an earlier version of the manuscript.
Finally, we thank Mohammad Akbar, Daniel Grumiller and Robert McNees for important questions and comments.  This
work was supported in part by NSERC (TL), a Frymoyer scholarship (DS), NSF grant PHY0854743, The George A. and
Margaret M. Downsbrough Endowment and the Eberly research funds of Penn State.


\end{document}